\documentclass[Physsubmission, Phys]{SciPost}

\hypersetup{
    colorlinks,
    linkcolor={red!50!black},
    citecolor={blue!50!black},
    urlcolor={blue!80!black}
}

\usepackage[bitstream-charter]{mathdesign}
\DeclareSymbolFont{usualmathcal}{OMS}{cmsy}{m}{n}
\DeclareSymbolFontAlphabet{\mathcal}{usualmathcal}

\begin{document}

\begin{center}{\Large \textbf{
Intrinsic charm in the nucleon and forward production of charm:\\ a new constrain from IceCube Neutrino Observatory
}}\end{center}

\begin{center}
Rafa{\l} Maciu{\l}a\textsuperscript{1$\star$},
Victor P. Goncalves\textsuperscript{2} and
Antoni Szczurek\textsuperscript{1,3}
\end{center}

\begin{center}
{\bf 1} Institute of Nuclear Physics, Polish Academy of Sciences, ul. Radzikowskiego 152, PL-31-342 Krak{\'o}w, Poland
\\
{\bf 2} Instituto de F\'{\i}sica e Matem\'atica,  Universidade Federal de Pelotas (UFPel), \\Caixa Postal 354, CEP 96010-900, Pelotas, RS, Brazil
\\
{\bf 3} University of Rzesz\'ow, PL-35-959 Rzesz\'ow, Poland
\\
* rafal.maciula@ifj.edu.pl
\end{center}

\begin{center}
\today
\end{center}


\definecolor{palegray}{gray}{0.95}
\begin{center}
\colorbox{palegray}{
  \begin{tabular}{rr}
  \begin{minipage}{0.1\textwidth}
    \includegraphics[width=22mm]{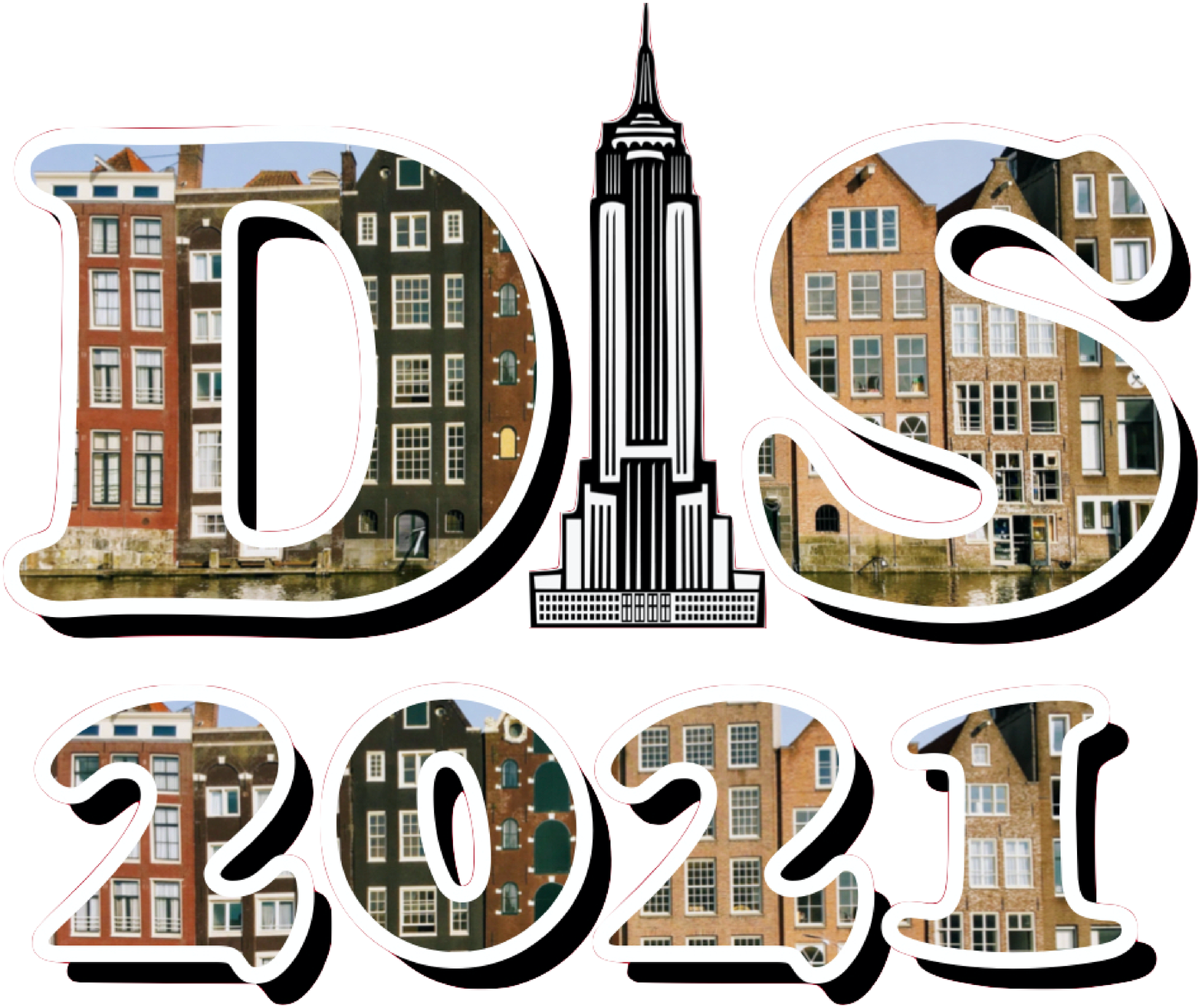}
  \end{minipage}
  &
  \begin{minipage}{0.75\textwidth}
    \begin{center}
    {\it Proceedings for the XXVIII International Workshop\\ on Deep-Inelastic Scattering and
Related Subjects,}\\
    {\it Stony Brook University, New York, USA, 12-16 April 2021} \\
    \doi{10.21468/SciPostPhysProc.?}\\
    \end{center}
  \end{minipage}
\end{tabular}
}
\end{center}

\section*{Abstract}
{\bf
The predictions for the atmospheric neutrino flux at high energies
strongly depend on the contribution of prompt neutrinos, which are
determined by the production of charmed meson in the atmosphere 
at very forward rapidities. Here we estimate the related 
cross sections taking into account the presence of an intrinsic charm (IC) component 
in the proton wave function. The impact on the predictions for the prompt neutrino flux is
investigated assuming different values for the probability to find the
IC in the nucleon.
}


\section{Introduction}
\label{sec:intro}

Recent experimental results obtained by the LHC, the Pierre
Auger and  IceCube Neutrino Observatories have challenged understanding of many interesting aspects of Quantum Chromodynamics in the high energy limit. In particular, in recent years, IceCube measured the astrophysical and atmospheric
neutrinos fluxes at high energies (see e.g. \cite{ice1}) while different collaborations
from the LHC performed several analyses of the heavy meson production at
high energies and forward rapidities (see e.g. \cite{LHCb:2015swx}).  
Those different origin data sets are strictly interrelated, since the
description of the heavy meson production  
at the LHC and higher center of mass energies is fundamental to make
precise predictions of the prompt neutrino flux \cite{Goncalves:2017lvq}, which
is expected to dominate the atmospheric $\nu$ flux for large neutrino energies
\cite{Ahlers:2018mkf}.

This aspect motivates the development
of new and/or more precise approaches to describe the perturbative and
nonperturbative regimes of the Quantum Chromodynamics (QCD) needed to
describe the charmed meson production in a kinematical range  beyond
that reached in hadronic collisions at the LHC. For this new kinematical
range, some topics are theme of intense debate. An important question, 
which motivates the present study, is whether the current and future
IceCube data can put some constraints on the intrinsic charm concept in the nucleon.

\section{Formalism}

The atmospheric neutrinos are produced in cosmic-ray interactions with nuclei
in Earth's atmosphere \cite{Ahlers:2018mkf}. While at low neutrino
energies ($E_{\nu}\lesssim 10^5$ GeV), these neutrinos arise from the
decay of light mesons (pions and kaons), and the associated flux is
denoted as the {\it conventional} atmospheric neutrino flux 
\cite{Honda:2006qj}, for larger energies  
it is expected that the {\it prompt} atmospheric neutrino flux associated with the
decay of hadrons containing heavy flavours become important
\cite{ingelman}.

Calculations of the prompt atmospheric neutrino flux at the
detector level depend on the description of the  production and decay of the
heavy hadrons as well as the propagation of the associated particles
through the atmosphere. Following our previous studies \cite{Goncalves:2017lvq,Goncalves:2018zzf}, we
estimated the expected prompt neutrino flux in the detector 
$\phi_{\nu}$  using the $Z$-moment method \cite{ingelman}, which implies
that  $\phi_{\nu}$  can be estimated using the geometric interpolation formula
\begin{eqnarray}
\phi_{\nu} = \sum_H \frac{\phi_{\nu}^{H,low} \cdot  \phi_{\nu}^{H,high}} {\phi_{\nu}^{H,low} + \phi_{\nu}^{H,high}} \,\,.
\label{eq:flux}
\end{eqnarray}
where $H = D^0, D^+, D_s^+$, $\Lambda_c$ for charmed hadrons and
${\phi_{\nu}^{H,low}}$ and ${\phi_{\nu}^{H,high}}$ are solutions of a
set of coupled cascade equations for the nucleons, heavy meson 
and lepton (and their antiparticles) 
fluxes in the low- and high-energy ranges, respectively. 
They can be expressed in terms of the nucleon-to-hadron
($Z_{NH}$), nucleon-to-nucleon ($Z_{NN}$), hadron-to-hadron  ($Z_{HH}$) 
and hadron-to-neutrino ($Z_{H\nu}$) $Z$-moments,  as follows \cite{ingelman}:
\begin{eqnarray}
\phi_{\nu}^{H,low} & = & \frac{Z_{NH}(E) \, Z_{H\nu}(E)}{1 - Z_{NN}(E)} \phi_N (E,0) \,, \\
\phi_{\nu}^{H,high} & = & \frac{Z_{NH}(E) \, Z_{H\nu}(E)}{1 - Z_{NN}(E)}\frac{\ln(\Lambda_H/\Lambda_N)}
{1 - \Lambda_N/\Lambda_H} \frac{m_H c h_0}{E \tau_H} f(\theta) \, \phi_N (E,0) \,,
\end{eqnarray}
where $\phi_N(E,0)$ is the primary 
flux of nucleons in the atmosphere, $m_H$ is the decaying particle's mass, $\tau_H$ is the proper 
lifetime of the hadron, $h_0 = 6.4$ km, 
$f(\theta) \approx 1/\cos \theta$ for $\theta < 60^o$, and the effective 
interaction lengths $\Lambda_i$ are given by $\Lambda_i = \lambda_i/(1 -
Z_{ii})$, with $\lambda_i$ being the associated interaction length ($i =
N,H$). For $Z_{H\nu}$, our treatment of the semileptonic decay of $D$-hadrons follows closely Ref. \cite{anna2}.
For a detailed discussion of the cascade equations, see
e.g.~Ref.~\cite{ingelman}. 
Assuming that the incident flux can be represented by protons ($N = p$), the charmed 
hadron $Z$-moments are given by
\begin{eqnarray}
Z_{pH} (E) =  \int_0^1 \frac{dx_F}{x_F} \frac{\phi_p(E/x_F)}{\phi_p(E)} 
\frac{1}{\sigma_{pA}(E)} \frac{d\sigma_{pA \rightarrow H}(E/x_F)}{dx_F} \,\,,
\label{eq:zpc}
\end{eqnarray}
where $E$ is the energy of the produced particle (charmed meson), $x_F$ 
is the Feynman variable, $\sigma_{pA}$ is the inelastic proton-Air
cross section and 
$d\sigma/dx_F$ is the differential cross section for the charmed meson 
production. 

As discussed in Ref.~\cite{antoni_rafal_jhep}, the cross section for
charm production at large forward rapidities, which is the region of
interest for estimating the prompt $\nu_{\mu}$ flux
\cite{Goncalves:2017lvq}, can be expressed as
\begin{eqnarray}
d \sigma_{pp \rightarrow charm} = d \sigma_{pp \rightarrow charm}(gg \rightarrow c \bar{c}) + d \sigma_{pp \rightarrow charm}(cg \rightarrow  cg) \,\,,
\label{eq:fac}
\end{eqnarray}
where the first and second terms represent the contributions associated
with the  $gg \rightarrow c \bar{c}$ and $ cg \rightarrow cg $ mechanisms,
with the  corresponding expressions depending  on the factorization
scheme assumed in the calculations. In
Ref.~\cite{antoni_rafal_jhep}, a detailed comparison between the
collinear, hybrid and $k_T$-factorization approaches was performed. 
In what follows, we will focus on the hybrid factorization model.
In this approach, the differential cross sections for $gg^* \rightarrow
c\bar{c}$ and $cg^* \rightarrow  cg$  mechanisms are given by
\begin{eqnarray}
d \sigma_{pp \rightarrow charm}(gg \rightarrow c \bar{c}) = \int dx_1
\int \frac{dx_2}{x_2} \int d^2k_t \, g(x_1,\mu^2) \, {\cal{F}}_{g^*}
(x_2, k_t^2, \mu^2) \, d\hat{\sigma}_{gg^* \rightarrow  c\bar{c}} \; ,
\label{eq:dif}
\end{eqnarray}
\vskip-5mm
\begin{eqnarray}
d \sigma_{pp \rightarrow charm}(cg \rightarrow  cg) = \int dx_1  \int \frac{dx_2}{x_2} \int d^2k_t \, c(x_1,\mu^2) \, {\cal{F}}_{g^*} (x_2, k_t^2, \mu^2) \, d\hat{\sigma}_{cg^* \rightarrow  cg} \,\,,
\label{eq:dif2}
\end{eqnarray}
where $g(x_1,\mu^2)$ and $c(x_1,\mu^2)$ are the collinear PDFs in the
projectile, ${\cal{F}}_{g^*} (x_2, k_t^2, \mu^2)$ is the unintegrated
gluon distribution (gluon uPDF) of the proton target, $\mu^2$ 
is the factorization scale of the hard process and the subprocesses 
cross sections are calculated assuming that the small-$x$ gluon is off 
mass shell and are obtained from a gauge invariant tree-level off-shell 
amplitude. In our calculations $c(x_1,\mu^2)$, similarly 
$\bar c(x_1,\mu^2)$, contain the intrinsic charm component.

As emphasized in Ref.~\cite{antoni_rafal_jhep}, the hybrid 
model, already at leading-order,
takes into account radiative higher-order corrections associated with
extra hard emissions that are resummed by the gluon uPDF.
In the numerical calculations below the intrinsic charm PDFs are taken at the initial scale $m_c=  1.3$ GeV,  so the perturbative charm contribution is intentionally  not  taken into account when discussing IC contributions.

Considering the $cg^* \rightarrow  cg$ mechanism one has to deal with
the massless partons (minijets) in the  final state. The relevant
formalism with massive partons is not yet available. Therefore it is
necessary to regularize the cross section that has a singularity in the
$p_{t} \rightarrow 0 $ limit. We follow here the known prescription
adopted in \textsc{Pythia}, where a special suppression factor is
introduced at the cross section level. The form factor depends on a free
parameter $p_{t0}$, which can be fixed using experimental
data for the $D$ meson production in $p + p$ and $p+^4He$ collisions at 
$\sqrt{s} = 38.7$ GeV and 86 GeV, respectively (see e.g. Ref.~\cite{Maciula:2021orz}). In numerical calculations below we use $p_{t0} = 2$ GeV.

The predictions for the charm production strongly depend on
modelling of the partonic content of the proton
~\cite{antoni_rafal_jhep}. In particular, the contribution of the charm
- initiated process is directly associated with the description of the
extrinsic and intrinsic components
(for a recent review see, e.g. Ref.~\cite{Brodsky:2020zdq}).  
Differently from the extrinsic charm quarks/antiquarks that 
are generated perturbatively by gluon splitting, the intrinsic 
one have multiple connections to  the valence quarks of the proton 
and thus is sensitive to its nonperturbative structure. 
The presence of an intrinsic component implies a large enhancement 
of the charm distribution  at large  $x$  ($> 0.1$) in comparison to 
the extrinsic charm prediction. In recent years, 
the presence of an intrinsic charm component  have been included in
the initial conditions of the global parton analysis \cite{ct14}, 
resulting in IC distributions that are compatible with the world
experimental data. 
However, its  existence is still a subject of intense debate, mainly associated with the amount of intrinsic
charm in the  proton wave function, which is directly related to 
the magnitude of the probability to find an intrinsic charm or anticharm
($P_{ic}$) in the nucleon.

In our analysis we will consider the collinear PDFs
given by the CT14nnloIC parametrization \cite{ct14} from
a global analysis assuming that the $x$-dependence of
the intrinsic charm component is described by the BHPS model
\cite{bhps}. Another important
ingredient is the modelling of ${\cal{F}}_{g^*} (x_2, k_t^2, \mu^2)$. In our analysis here
we will use the uPDF derived using the Kimber-Martin-Ryskin (KMR) prescription \cite{kmr}.

\section{Numerical results}
\label{sec:results}

\begin{figure}[h]
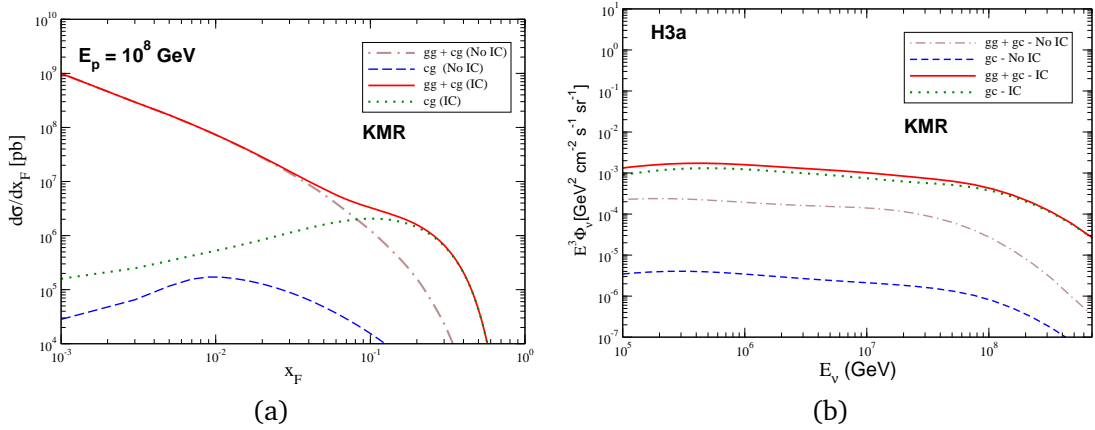

\begin{center}
\begin{tabular}{cc}
\includegraphics[width=0.45\textwidth]{dsigdxf_icxnoic_KMR_pt2.eps} &  
\includegraphics[width=0.45\textwidth]{fluxE3_KMR_pt2.eps} \\
(a) & (b)
\end{tabular}

\end{center}
\caption{Predictions of the hybrid model for (a) the Feynman $x_F$ -
  distributions for charm particles and (b) the prompt neutrino flux
  (rescaled by $E_{\nu}^3$).}
\label{Fig:kmr}
\end{figure} 

In Fig.~\ref{Fig:kmr} (a), we present our predictions for the
Feynman $x_F$ distribution of charm particles produced in $pp$
collisions at the atmosphere, considering an incident proton with an
energy of $E_p = 10^8$ GeV and the KMR model for the uPDF. 
We present separately the contribution associated with the $cg
\rightarrow cg$ mechanism and the sum of the two mechanisms, denoted  
by ``cg'' and  ``gg + cg'', respectively. 
Moreover, we compare the IC predictions, obtained using the CT14nnloIC
parametrization for $P_{ic} = 1\%$, with those obtained disregarding 
the presence of the intrinsic component (denoted No IC hereafter). 
One has that for small $x_F (\equiv  x_1 - x_2)$, the charm production
is dominated by the $gg \rightarrow c \bar{c}$ mechanism, which is
expected since for  $x_F   \approx 0$ and high energies both
longitudinal momentum fractions $x_i$ are very small and the proton
structure is dominated by gluons. For the No IC case, the  contribution
of the $cg \rightarrow cg$ mechanism is smaller than the gluon fusion one
for all values of $x_F$. In contrast, when intrinsic charm is included,
the behavior of the distribution in the intermediate $x_F$ range ($0.06
\le x_F \le 0.6$) is strongly modified. Such a behaviour is expected,
since for this kinematical range, the charm production depends on
the description of the partonic content of the incident proton at large
values of the Bjorken $x$ variable. 
The impact on the predictions for the prompt neutrino flux is
presented in Fig.~\ref{Fig:kmr} (b). As expected from the analysis
performed in Ref.~\cite{Goncalves:2017lvq}, where we find that the
dominant contribution to the neutrino flux comes typically from $x_F$ in
the region $0.2<x_F<0.5$, one has that the flux is enhanced  by one
order of magnitude when intrinsic charm is included. In agreement with
the results presented in Fig.~\ref{Fig:kmr} (a), the contribution of the
$cg \rightarrow cg$ mechanism is negligible for the No IC case. 
However, it becomes dominant in the IC case, with the normalization of
the prompt flux dependent on the amount of IC.

\begin{figure}[t]
\begin{center}
\includegraphics[width=0.45\textwidth]{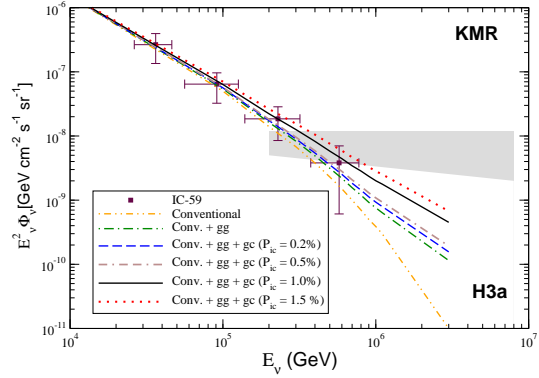}  
\end{center}
\caption{Comparison between our predictions and the experimental IceCube
  data \cite{ice1} for the atmospheric $\nu_{\mu}$ flux for the
  KMR uPDFs.}
\label{Fig:fluxe2}
\end{figure} 

In Fig.~\ref{Fig:fluxe2} we present our results for the atmospheric
$\nu_{\mu}$ flux, scaled by a factor $E_{\nu}^2$, which is the sum of
the conventional and prompt contributions. The predictions were obtained
considering different values for
$P_{ic}$ in the calculation of the prompt contribution. Moreover, for
the conventional  atmospheric neutrino flux we assume the result derived
in Ref.~\cite{Honda:2006qj}. The resulting predictions are compared with
the IceCube data obtained in Ref.~\cite{ice1} for the zenith-averaged flux of atmospheric neutrinos.
 One has that the prompt contribution enhances the flux at large
 neutrino energies, with the enhancement being strongly dependent on the
 magnitude of the $cg \rightarrow cg$ mechanism. If this mechanism is disregarded, the results
 represented by  ``Conv. + gg'' in the figures indicate that the impact
 of the prompt flux is small in the current kinematical range probed by
 IceCube. 
On the other hand, the inclusion of the $cg \rightarrow cg$ mechanism
implies a large enhancement of the prompt flux at large $E_{\nu}$, with
the associated magnitude being strongly dependent on the value of
$P_{ic}$. Our results for the KMR uPDF, presented in Fig.~\ref{Fig:fluxe2}, indicate that a value of $P_{ic}$ larger than $1.5 \%$ implies a prediction
for neutrino flux that overestimate the IceCube data at high energies.
This result sets the upper limit for the intrinsic charm ammount in the nucleon to $P_{ic} \leq 1.5\%$. 
Surely, future data can be more
restrictive in the acceptable range of values for $P_{ic}$.  It was also
shown in our original paper \cite{Goncalves:2021yvw} that presence of saturation effects leads to a slightly larger values for $P_{ic}$ that are not discarded by the current IceCube data.

\vspace{-5mm}

\section{Conclusion}
We have investigated the impact of the intrinsic charm 
component in the hadron wave function, 
which carries a large fraction of the hadron momentum, on the prompt neutrino flux.
Our results has indicated that the inclusion of the $cg \rightarrow cg$ mechanism has a strong effect 
on the prompt neutrino flux. In particular, when the IC component 
is present, such a mechanism determines the energy dependence of 
the flux at high energies, with the normalization dependent on 
the value assumed for the probability to find the IC in 
the proton wave function.  

\vspace{-5mm}

\section*{Acknowledgements}

\vspace{-3mm}

This study was supported by the Polish National Science Center
grant UMO-2018/31/B/ST2/03537 and by the Center for Innovation and
Transfer of Natural Sciences and Engineering Knowledge in Rzesz{\'o}w
and by the Brazilian funding agencies CNPq, FAPERGS and INCT-FNA
(process number 464898/2014-5).

\end{document}